\documentclass[prb,reprint,showpacs,amsmath,amssymb,superscriptaddress,citeautoscript,longbibliography]{revtex4-1}

\usepackage[plainpages=false,pdfpagelabels,colorlinks=true,linkcolor=red,urlcolor=blue,citecolor=blue,pdftitle={Title},pdfauthor={},pdfdisplaydoctitle=true,pdfduplex=DuplexFlipLongEdge]{hyperref}
\usepackage{epsfig}
\usepackage{amsmath,amssymb}
\usepackage{physics}
\usepackage{graphicx}
\usepackage[dvipsnames,usenames]{color}
\usepackage[normalem]{ulem}
\usepackage{soul}

\begin{document}

\title{Enhancement of Charge Density Wave Correlations in a Holstein 
Model 
with an Anharmonic Phonon Potential}

\author{C. Kvande}
\email[Corresponding author: ]{Claire.Kvande19@kzoo.edu}
\affiliation{Physics Department,
Kalamazoo College, 1200 Academy Street, Kalamazoo, Michigan,
49006-3295 USA}
\affiliation{Department of Physics, University of California, Davis,
  California 95616, USA}  
\author{C. Feng}
\affiliation{Center for Computational Quantum Physics,
Flatiron Institute, 162 Fifth Avenue, New York, New York 10010} 
\affiliation{Department of Physics, University of California, Davis,
  California 95616, USA}
\author{F. H\'ebert}
\affiliation{Universit\'e C\^ote d'Azur, CNRS, INPHYNI, France}
\author{G.$\,$G. Batrouni}
\affiliation{Department of Physics, National University of Singapore,
  2 Science Drive 3, 117551 Singapore} 
\affiliation{Centre for Quantum Technologies, National University of
  Singapore; 2 Science Drive 3, 117543 Singapore} 
\affiliation{Universit\'e C\^ote d'Azur, CNRS, INPHYNI, France}
\author{R.$\,$T. Scalettar}
\affiliation{Department of Physics, University of California, Davis,
  California 95616, USA}

\begin{abstract}
The Holstein Hamiltonian describes itinerant electrons whose site
density couples to local phonon degrees of freedom.  In the single site
limit, at half-filling, the electron-phonon coupling results in a double
well structure for the lattice displacement, favoring empty or doubly
occupied sites.  In two dimensions, and on a bipartite lattice in $d \geq 2$, an intersite hopping causes these
doubly occupied and empty sites to alternate in
a charge density wave (CDW) pattern when the temperature is lowered.  Because a discrete symmetry
is broken, this occurs in a conventional second-order  transition at finite $T_{\rm cdw}$.  In this
paper, we investigate the effect of changing the phonon potential energy
to one with an {\it intrinsic}  double well structure even in the absence
of an electron-phonon coupling.  While this aids in the initial process
of pair formation, the implications for subsequent CDW order are
non-trivial.  One expects that, when the electron-phonon coupling is too
strong, the double wells become deep and the polaron mass large, an
effect which {\it reduces} $T_{\rm cdw}$.   We show here the existence
of regions of parameter space where the double well potential, while
aiding local pair formation, does so in a way which also substantially enhances long range CDW order.
\end{abstract}


\maketitle

\section{Introduction}

The Holstein Hamiltonian\cite{holstein1959} provides a simplified
description of the interactions between electron and phonon degrees of
freedom in a solid, including polaron and bipolaron formation
\cite{kornilovitch98,kornilovitch99,alexandrov00,hohenadler04,ku02,spencer05,macridin04,romero99,bonca99},
and the origin of low temperature phases with diagonal charge density
wave (CDW) or off-diagonal superconducting (SC) long range order
\cite{peierls79,hirsch82,hirsch83,scalettar89,marsiglio90,freericks93,ohgoe17,hohenadler19,bradley21,nosarzewski21,araujo22}.
Although the electron-phonon interaction, $\lambda$, initiates these
phases, its effect is non-monotonic\cite{blawid00,weber2018,zhang2019,feng2020,feng2020i,chen2019,cohenstead20,bradley21}.   
Quantum Monte Carlo
(QMC) simulations show that pairs become heavy and CDW and SC transition
temperatures go to zero at strong coupling $\lambda$\cite{esterlis18}.
This finding is in contrast with the approximate Eliashberg theory, which
predicts that $T_{\rm cdw}$ increases monotonically with $\lambda$, and provides a challenge to
achieving high CDW transition temperatures.

As a consequence, the search for situations in which large $\lambda$
does not reduce the tendency for long range order has been
an ongoing focus of recent analytic and numerical studies.
For example, in the case of SC, it has been suggested 
that a Su-Schrieffer-Heeger (SSH)
interaction \cite{xing2021,feng2022,assaad2022,cai2021} might avoid the problem of large effective
mass\cite{sous18,sous23}.  Elevated CDW transitions have also been
found in studies of the SSH model on a 3D Lieb lattice appropriate to
the bismuthates\cite{cohenstead22bismuth}.

In infinite dimension, using a technique similar to dynamical mean field theory
(DMFT), Freericks, Jarrell, and Mahan \cite{freericks96} studied
the effects of a simple anharmonic term in the form of an additional 
quartic potential energy for the phonons. 
They concluded that a CDW phase exists for a large range of densities at low
anharmonicity, but that the CDW is gradually replaced at low and high densities
by a SC phase as the anharmonicity increases. The half-filled system
always remains in a CDW state. They also observed a decrease of the
critical temperatures at which CDW and SC phases appear with
increasing anharmonicity. Similar models have been studied in
one dimension \cite{chatterjee04}.

In this manuscript we consider a route to high CDW transition
temperatures driven by a double well (anharmonic) phonon potential
resulting from {\it negative} quadratic, and positive quartic, terms in the
displacement.  
Such a potential favors the development of a preexisting non zero phonon field, without the mediation of electron-phonon coupling, and then favors electron occupations to organize into empty and doubly occupied sites when the el-ph interaction is present. A number of previous studies of anharmonicity with 
{\it positive} quadratic and positive quartic phonon potential energy terms \cite{chatterjee04,adolphs2013,li2015-1,li2015-2,dee20,uma17,freericks96,hirsch83,hui74,kavakozov78,mahan93,szabo21},
have in general found a suppression of charge order at half-filling, 
in agreement with
the DMFT study noted above. Nonlinearities in the coupling terms between
fermions and phonons \cite{adolphs2013,li2015-1,li2015-2,dee20} have led to similar conclusions.  This existing
literature brings into focus our key
result: anharmonicity can produce an enhancement of 
$T_{\rm cdw}$ if it occurs in the form of an intrinsic double well
potential.

There are a number of experimental motivations for considering
such a generalization of the Holstein Hamiltonian.
One is to understand Kondo/heavy fermion physics 
in materials like SmOs$_4$Sb$_{12}$.  Most typically, 
heavy fermion behavior arises due to the interaction of
conduction electrons with {\it magnetic} degrees of freedom (local moments).
However, it has been suggested, even dating back to Kondo\cite{kondo76}, 
that other two level systems might cause similar phenomena.
In the case of  SmOs$_4$Sb$_{12}$ a large applied magnetic field, 
which would quench fluctuations of local magnetic moments and
hence of Kondo physics, does not destroy the heavy fermion
behavior.  It has been suggested, then, that rather than
conduction electrons interacting
with local $S=1/2$ spins, it is instead the coupling
to two level phonon degrees of freedom that is relevant\cite{fuse11,fuse12}.

The remainder of this paper is organized as follows:
We first define the model, its parameters and physical observables,
and then give a brief summary of our two, complementary, QMC
techniques.  Results are then shown for local observables
and for charge structure factors
for different forms of the anharmonic potential using
energy scales close to those typically chosen in the 
conventional Holstein model.  Finite size scaling (FSS)
is employed to extract $T_{\rm cdw}$.
Similar calculations are then done for parameters
which fix the average phonon displacement in order to demonstrate that the enhanced CDW $T_{\rm cdw}$
is not a `trivial' effect associated with artificially large displacements.
A conclusion summarizes our work and points
to possible future directions, and is followed by
Appendices containing further details of our model and simulations.

\section{Model and methods}

We consider the Hamiltonian,
\begin{eqnarray}
H &=&-t \sum_{\langle ij\rangle\sigma}
\big(c^\dagger_{i\sigma}c^{\phantom{\dagger}}_{j\sigma}+ h.c.\big)
-\mu \sum_{i\sigma} n_{i\sigma}
\nonumber \\
&&+\sum_i \big(-A x_i^2 +
B x_i^4 +
\frac{p_i^2}{2m}\big) 
\nonumber \\
&&+
\lambda \sum_{i} x_i \, \big( n_{i\uparrow}+n_{i\downarrow} -1 \big)
\label{eq:ham}
\end{eqnarray}
The sums run over the $N=L^2$ sites of a two-dimensional square lattice.
The operator $c^{\phantom\dagger}_{i\sigma}$ ($c^\dagger_{i\sigma}$)
destroys (creates) a fermion of spin $\sigma =\,\uparrow {\rm or}
\downarrow$ on site $i$; $n^{\phantom{\dagger}}_{i\sigma} = c_{i\sigma}^\dagger
c^{\phantom\dagger}_{i\sigma}$ is the corresponding number operator;
$x_i$ and $p_i$ are the canonically conjugate displacement and momentum operators
of the phonon mode at site $i$.  The first line of Eq.~\ref{eq:ham}
represents the hopping energy of the fermions between neighboring
sites $\langle ij\rangle$.  A chemical potential term is included to
emphasize
our algorithms perform simulations in the grand canonical
ensemble.  The hopping parameter $t$ will be used as the energy scale.
The second line in Eq.~\ref{eq:ham} represents the energy of the
phonons of 
quadratic 
potential $-Ax_i^2$ and anharmonic term
$B x_i^4$.  
This form, with a negative quadratic term (i.e.~$A>0$),
results in a double well.
Without loss of generality, we set $m=1$.
The third line in Eq.~\ref{eq:ham} is the
phonon-electron interaction, written in a particle hole symmetric (PHS) form 
so that $\mu=0$ corresponds to half-filling.  
A further discussion of this PHS appears in Appendix 1.
The PHS also ensures the values of displacement $x$ corresponding to
empty and doubly occupied sites are symmetrically
located about the origin $x=0$.

In order to connect to previous QMC studies of
the conventional Holstein Hamiltonian\cite{holstein1959}, where there is
only a positive quadratic term in the phonon displacement with phonon frequency $\omega_0$,
we note that one would express the quadratic 
coefficient in terms of the frequency, as
$A = m\omega_0^2/2$.  
In that situation, $\omega_0$ also 
enters the re-writing of the electron-phonon interaction
in terms of phonon creation (destruction) operators,
$a^\dagger_i (a_i)$:
$
\lambda \sum_{i} x_i \, \big(n_{i\uparrow}+n_{i\downarrow} -1\big)
= 
g \sum_{i} \big( a^{\phantom{\dagger}}_i + a^{\dagger}_i \big)
\, \big(n_{i\uparrow}+n_{i\downarrow} -1\big)
$
with $g=\lambda/\sqrt{2\omega_0}$ where 
$\omega_0 = \sqrt{2A}$. 
To compare to previous work on the conventional 
Holstein model, we then choose a commonly used value of coupling $g$, keep $B$ fixed to a small value and vary $A$ to explore different depths of the potential wells.  The values of $A$ are chosen to keep $\omega_0=\sqrt{2A}$ and $\lambda=g\sqrt{2\omega_0}$ of order unity, in the range of values that are typically used for the conventional Holstein model. Results
corresponding to this choice of parameters will be presented in Sec. III.

However, although analogous values of the el-ph coupling
and phonon frequency are used in this comparison, the anharmonic form
of the full phonon potential leads to displacements which are different
in magnitude from the simplest harmonic situation.  One can ensure that the
coupling to the electrons, which combines $\lambda$ and $x_i$,
is equivalent in magnitude to the conventional Holstein case
by choosing parameters $A$ and $B$ which are tuned to keep the average phonon displacement fixed at a certain value
$x_0$, where  $x_0$ is given by $\lambda/\omega_0^2$ in the conventional Holstein case.
This is accomplished through the choice
$A = {(4Bx_0^3 - \lambda)}/{(2 x_0)}$,
a relation derived in Appendix 2;
results corresponding to this choice of
parameters will be presented in Sec. IV.

We employ two methods to study Eq.~\ref{eq:ham}.  The
first is Determinant Quantum Monte Carlo (DQMC)\cite{blankenbecler81}.
In this approach, the action for the phonon degrees of freedom 
at inverse temperature (imaginary time) $\beta$ is
expressed as a path-integral over a space-imaginary time grid,
and the fermionic degrees of freedom, which appear
only quadratically in Eq.~\ref{eq:ham}, are integrated out analytically.
The resulting partition function consists of an integral
over the phonon field $x_i(\tau)$ which is performed 
stochastically.  The weight for phonon field configurations
takes the form of the square of the determinant of a matrix 
(the fermionic traces over spin up and down yield identical
determinants) whose dimension is the number of spatial sites $N$.
Consequently, there is no sign problem.  However, a sweep through the space-time lattice scales as $N^3 \beta$, and possibly as $N^3 \beta^2$, depending on the degree to which numerical instabilities require more accurate (numerically stable) treatment of the linear algebra.

DQMC studies of the conventional Holstein model date
back to the same period as for the Hubbard model\cite{scalettar89,noack93,vekic92,freericks93,marsiglio93}
but precise quantitative values for $T_{\rm cdw}$ have emerged
only more recently e.g. on square
\cite{weber2018}, honeycomb\cite{zhang2019} and
cubic lattices\cite{cohenstead20}.
The delay in computing the transition temperature partly originated in the 
quantum simulation 
community's focus instead on electron-electron
interactions as driving exotic superconductivity in the cuprates,
but also because of the significant computational challenge of very long autocorrelation times.
In DQMC simulations of the Hubbard model, updates of
the Hubbard-Stratonivich field at a single space-time point
decorrelate very rapidly (a few sweeps of the lattice).
However, in DQMC for the Holstein model, autocorrelation times
are instead often hundreds or thousands of sweeps.

This bottleneck has led to the development of QMC
methods for electron-phonon Hamiltonians 
based on a Langevin update of the entire space-time
lattice\cite{batrouni19,cohenstead20,paleari21,cohenstead22algo,zhang22}.  
Such approaches can be formulated in a way which
scales linearly in $N$ (albeit with a smaller step size for
each move than in DQMC) via the replacement of the determinant
by an integration over a pseudofermion field.  Equally important
to this linear scaling, Fourier acceleration methods\cite{batrouni85,batrouni19}
can be employed to reduce autocorrelation times dramatically.
Alternate methods to address
long autocorrelation times use
machine-learning approaches
\cite{chen18} and
Wang-Landau sampling
\cite{yao21}.

We will employ both DQMC and Langevin methods here. Most of the simulations have been performed with DQMC and the results presented here were obtained with this method unless otherwise indicated in the figures. In certain key cases, results have been confirmed by comparing DQMC and Langevin simulations.

\begin{figure}[t!]
\centerline{\includegraphics[width=0.5\textwidth]{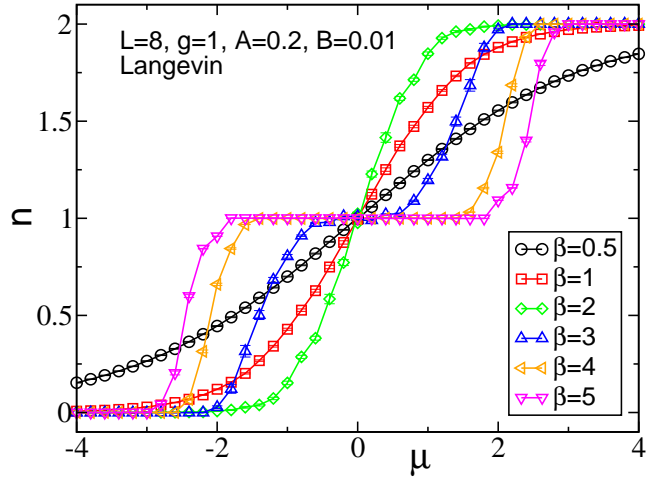}}
\caption{\label{fig:rhovsmu1}(Color online).
Density, $n$, as a function of chemical potential, $\mu$, for $g=1$, $B=0.01$ and $A=0.2$.  $\lambda=\sqrt{2\omega_0}g$ with $\omega_0=\sqrt{2A}$. At high temperature $T$, $n$ deviates immediately from half-filling as $\mu$ is changed from $\mu=0$.  However, as $T$ decreases a plateau in $n(\mu)$ develops:  the density is frozen at half-filling until $|\mu|$ exceeds a critical threshold, half the single particle gap $\Delta$. This gap formation around $\beta \simeq 3$ is an indication of the entry into the ordered CDW phase at low $T$. The simulations were performed only for $\mu\ge 0$ since the system is particle-hole symmetric.
 }
\end{figure}

The most simple observable we study is 
the density, $n=\sum_i\langle n_{i\sigma}\rangle/L^2$, and its behavior as a function of $\mu$. A plateau in $n(\mu)$ signals a vanishing compressibility, $\kappa = \partial n / \partial  \mu$,
and  the presence of a charge gap $\Delta$.  
As noted earlier, the PHS form of the Hamiltonian
ensures half-filling $n=1$ corresponds to $\mu=0$.
This is the optimal density for a CDW phase, since it allows a
precise alternation of doubly occupied and empty sites.
  
We will also examine other 
local quantities such as the average value of the phonon displacement
$\langle x_i\rangle$, the double occupancy $D=\langle
n_{i\uparrow}n_{i\downarrow}\rangle$, and the
kinetic energy $K\left\langle c^\dagger_{i,\sigma} c_{i+x,\sigma} +h.c. \right\rangle$.

To characterize further the presence of a (long range) CDW phase,
we study the charge structure factor,
the Fourier transform at momentum $(\pi,\pi)$ of
the density-density correlation function,
\begin{equation}
S_{\rm cdw} = \frac{1}{N}\sum_{i,j} \, \langle  n_i n_{i+j}\rangle (-1)^j \,\,.
\label{eq:Scdw}
\end{equation}
Here 
$n_i = n_{i\uparrow} + n_{i\downarrow}$ is the number of fermions on site $i$.
In a phase with short range order,
$\langle  n_i n_j\rangle $
will decay rapidly to zero as the separation $|i-j|$ increases.  Thus
in the sum over all pairs of sites $i,j$ in 
Eq.~\ref{eq:Scdw}, only sites $j$ in the immediate neighborhood of $i$ contribute, and 
the double sum is only of order $N$.  The division by $N$ then
implies $S_{\rm cdw} \sim o(1)$.  In a phase with long range order, on the other hand,
the double sum over pairs of sites is $o(N^2)$ and
$S_{\rm cdw} \sim o(N)$ after normalization.
The optimal ordering vector for a half-filled square lattice is 
at $(\pi,\pi)$ owing to the perfect nesting at this momentum.
Incommensurate order at $q \neq (\pi,\pi)$
is possible upon doping, but we do not see evidence of it here.

\begin{figure}[t!]
\centerline{\includegraphics[width=0.5\textwidth]{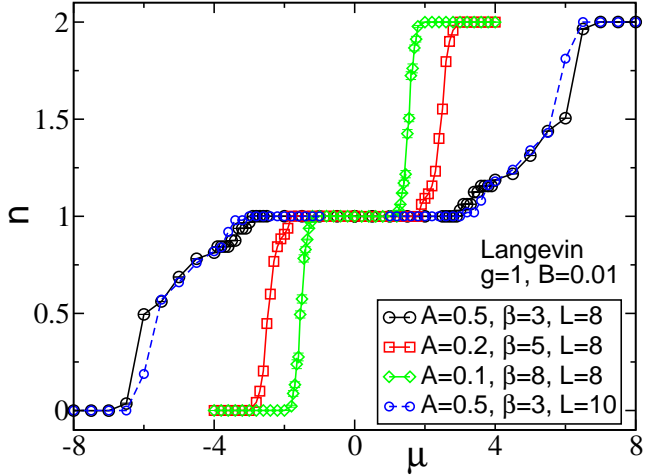}}
\caption{\label{fig:rhovsmu2}(Color online). Density, $n$, as
a function of chemical potential, $\mu$, for $g=1$, $B=0.01$ and several values of $A$ with  $\lambda=\sqrt{2\omega_0}g$ and $\omega_0=\sqrt{2A}$. $\beta$ is chosen
so that $n(\mu)$ no longer changes with further lowering of the temperature allowing the simulation to pick up only ground state properties.
(See also Fig.~\ref{fig:Svsbetaall}.) We observe a decrease of the charge gap as $A$ decreases from $0.5$ to $0.1$.
For the $A=0.5$ case, a comparison of results 
for $L=8$ and $L=10$ shows that the width of the gap does not change significantly with size.}
\end{figure}

\begin{figure}[t!]
\centerline{\includegraphics[width=0.5\textwidth]{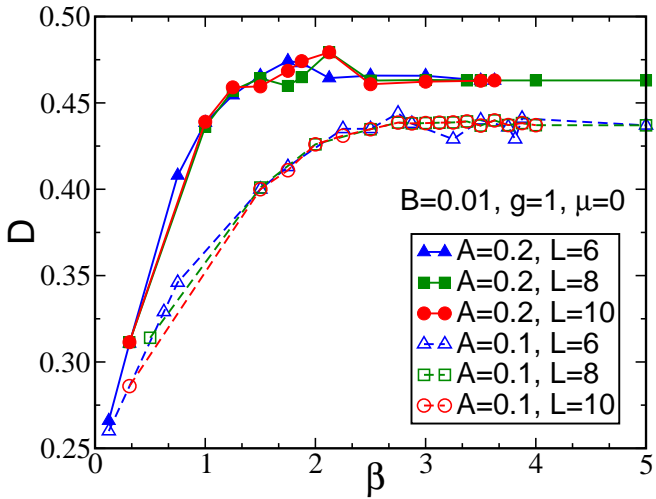}}
\caption{\label{fig:nudvsbeta}(Color online). Variation of the double occupancy $D$ with $\beta$ for $B=0.01$, $\mu = 0$, $g = 1$, for different sizes $L$, and for two different choices of $A$: $A=0.1$ (open symbols, dashed lines) and $A=0.2$ (filled symbols, solid lines).  $\lambda=\sqrt{2\omega_0}g$ with $\omega_0=\sqrt{2A}$. $D$ saturates to a larger value and at a higher temperature for $A=0.2$, compared to $A=0.1$.
}
\end{figure}

\section{Simulations for Canonical Parameters}

We first show results for values of Hamiltonian parameters similar
to those used in past studies of the
conventional Holstein Hamiltonian in order to facilitate comparison
of our results with those in the literature.
Specifically, we fix the electron-phonon coupling 
$ g \sum_{i} \big( a^{\phantom{\dagger}}_i + a^{\dagger}_i \big)
\, \big(n_{i\uparrow}+n_{i\downarrow} -1\big)$
at $g=1$,
and pick $A=0.1, 0.2, 0.5$.  These correspond to quadratic potential curvatures
$\omega_0^2/2$ with $\omega_0=\sqrt{2A}=0.44, 0.63, 1.00$, similar to the commonly used values 
$\omega_0 =0.5\,$-$\,2.0$
\cite{blawid00,li2015-1,li2015-2,weber2018,zhang2019,feng2020,feng2020i,chen2019,cohenstead20,bradley21}.   
When expressed in terms of a
coupling of fermionic density to lattice displacement,
$\lambda= \sqrt{2 \omega_0} \, g = 0.94, 1.12, 1.41$,
again in the usual range of $\lambda \sim 1$.

\subsection{Local Observables}

Phases with long range order are typically characterized by gaps in their single particle
excitation spectra.  As noted earlier, such gaps are most simply revealed via a vanishing of the
compressibility $\kappa = \partial n / \partial \mu$, i.e.~by a plateau in a plot of $n$ versus 
$\mu$.  In Fig.~\ref{fig:rhovsmu1} we fix $A=0.2$, $B=0.01$, $g=1$.
At high temperatures the compressibility at half-filling
($\mu=0$) is finite.  However, when $T \lesssim t/3$ the slope of $n(\mu=0)$
becomes small.  At $T=t/5$, $n$ remains fixed at $n \sim 1$ 
until $\mu$ exceeds $\mu \sim 2t$, indicating a CDW gap $\Delta \sim 4t$.
The non-monotonic evolution of  the compressibility in 
Fig.~\ref{fig:rhovsmu1} 
can be understood by the fact that, in addition to the non-trivial
physics of CDW formation which causes $\kappa \sim 0$ at low $T$, in  the limit of very high temperature the 
compressibility must also become small, ie $\kappa \sim 1/T$.

Figure \ref{fig:rhovsmu2} generalizes Fig.~\ref{fig:rhovsmu1} to several distinct values of $A$. As explained before, for each $A$, the electron-phonon coupling is chosen to mimic the procedure in the usual Holstein model, namely by identifying the frequency corresponding to the curvature, $\omega_0 = \sqrt{2A}$, and then determining the electron-phonon coupling
$\lambda = \sqrt{2\omega_0} \, g$ with $g$ fixed at $g=1$. 
Figure \ref{fig:rhovsmu2} allows us to assess how the single particle gap $\Delta$
is affected by the (negative) quadratic phonon curvature.
We find that $\Delta$ increases with increasing $A$.
We will return to this point in discussing the effect of 
varying $A$ on the CDW transition temperature.

We comment that for $A=0.5$, one can see additional steps in $n$ above half-filling. For $L=8$ ($N=64$), these occur at at integer densities corresponding to even numbers of particles  $N_\uparrow + N_\downarrow = 66,68, \cdots$ on the lattice and reflect the tendency to add particles in $\uparrow \downarrow$ pairs due to the attractive interaction mediated by the phonons. Similar steps are evident for $L=10$. This is an effect seen also in QMC simulations of the conventional Holstein model.

\begin{figure}[t!]
\centerline{\includegraphics[width=0.5\textwidth]{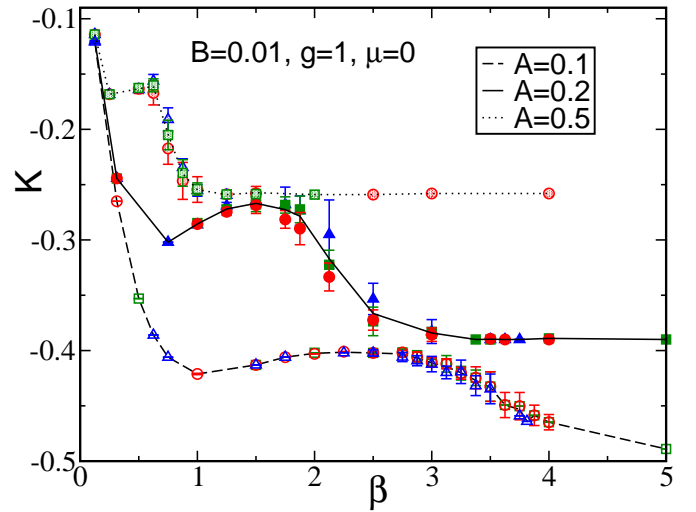}}
\caption{\label{fig:KExvsbeta}(Color online). $K$, the $x$ component of kinetic energy, as a function of inverse temperature for different values of $A=0.1, 0.2$ and 0.5 and different sizes $L=6,8$ and 10.
All data have $B=0.01$, $\mu=0$, $g=1$. $\lambda=\sqrt{2\omega_0}g$ with $\omega_0=\sqrt{2A}$.
Red circles are $L=10$, 
green squares $L=8$, and blue triangles are $L=6$. The lines show the average over the different lattice sizes at each point.
Because of particle-hole symmetry, the high temperature (small $\beta$) value of $K$ vanishes: The non interacting energy levels
$\epsilon({\bf k})$ are symmetric around $\epsilon=0$ and, at high $T$, all levels are occupied equally. As $\beta$ increases, the $\epsilon<0$ states are preferentially occupied, and $K<0$.
}
\end{figure}

The double occupancy $D $ is given in 
Fig.~\ref{fig:nudvsbeta} 
for two values of $A$ and different lattice sizes 
$L= 6, 8, 10$.
$D$ evolves rapidly from its high temperature (uncorrelated) value
$D = \langle n_\uparrow n_\downarrow \rangle \sim
\langle n_\uparrow \rangle \, \langle n_\downarrow \rangle \sim 1/4$
as $T$ decreases, reflecting the fact that pair formation precedes
the ordering of pairs into a CDW pattern.
The weak feature in $D$ at $\beta \sim 2$ 
will be seen to coincide with CDW formation.

A final local observable is the kinetic energy $K$, given in 
Fig.~\ref{fig:KExvsbeta}.  $K$ first evolves from its particle-hole symmetric
high temperature limit $K=0$, to negative values as lower energy states dominate.
This steady decrease is interrupted by upturns in $K$ (decreases in the
magnitude of hopping).  These local maxima correlate with the CDW ordering
transitions.  See below.

\begin{figure}[t!]
\centerline{\includegraphics[width=0.5\textwidth]{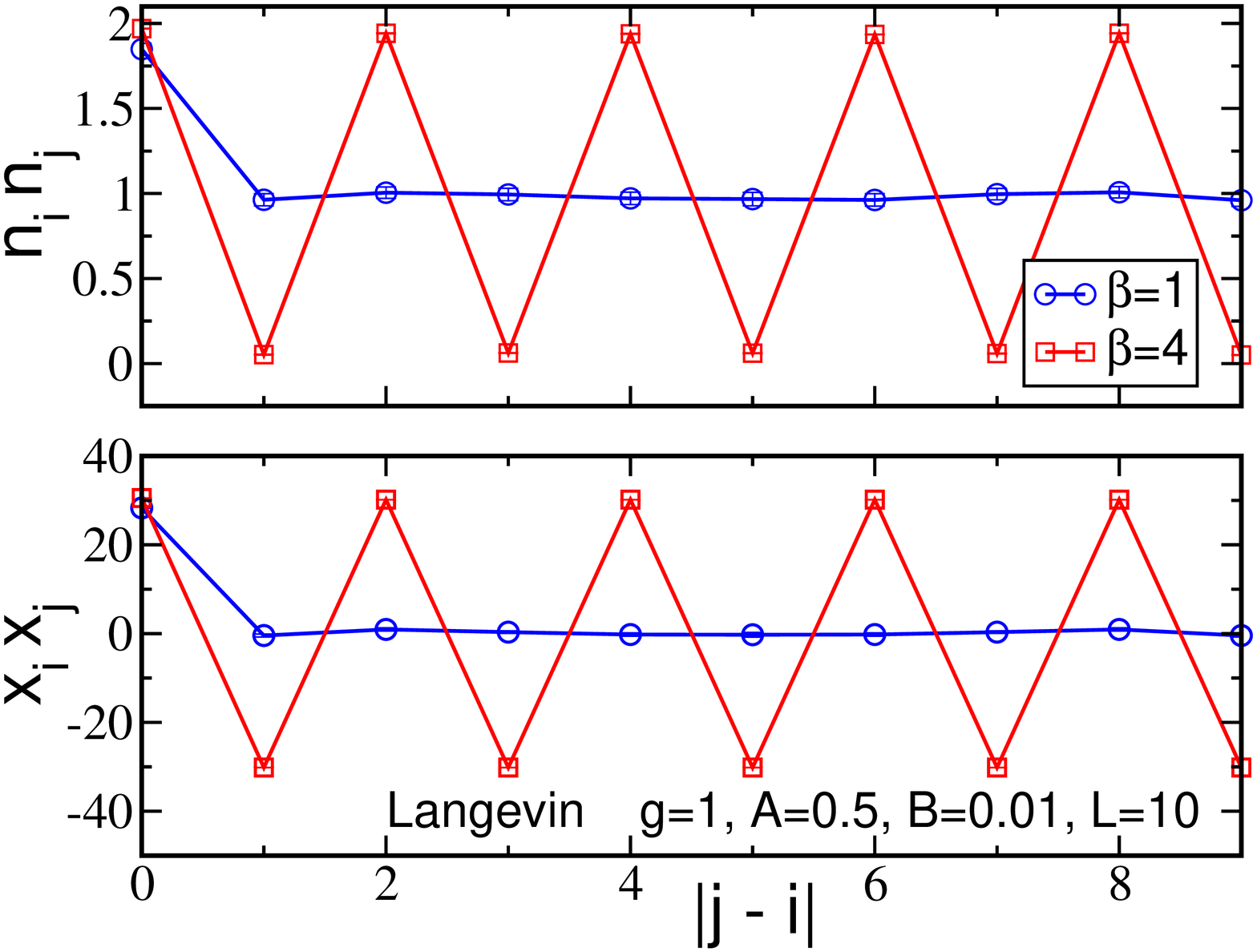}}
\caption{\label{fig:nnandxx}
(Color online). Density-density $\langle n_i n_j \rangle$ and phonon displacement correlations 
$\langle x_i x_j \rangle$ correlators along the side of the square lattice 
at high ($\beta=1$) and low ($\beta=4$) temperatures.
As $\beta$ increases the system goes from an unordered phase to 
a charge density wave phase where an order develop in both the charge density 
and phonon displacements. One should notice that, due to the double well
potential, $\langle x_i x_{j=i} \rangle$ is sizeable even in the high
temperature phase.
}
\end{figure}

\subsection{Long Range Charge Order}

\begin{figure}[t!]
\centerline{\includegraphics[width=0.5\textwidth]{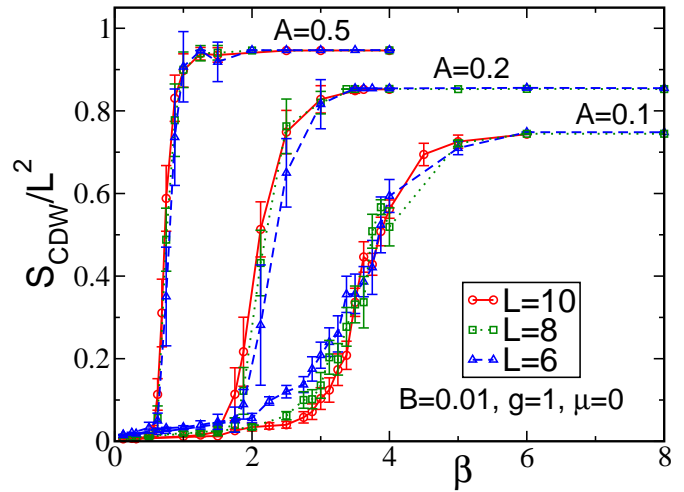}}
\caption{\label{fig:Svsbetaall}
(Color online).  Evolution of the charge structure factor $S_{\rm cdw}$ with
inverse temperature.  Here $B=0.01$, $\mu=0$, and $g=1$. From left to right, we have 
$A= 0.5$, $A= 0.2$, and $A= 0.1$ with corresponding
$\omega_0=\sqrt{2A}$ and
 $\lambda=\sqrt{2\omega_0}g$. As $A$ increases, the structure 
 factor $S_{\rm cdw}$ saturates at a larger value and the transition occurs
 at a larger temperature.
}
\end{figure}

Two final observables directly probe charge order.
The first, shown in the top panel of 
Fig.~\ref{fig:nnandxx},
is the real space density-density correlation function
$\langle n_i n_j \rangle$.  
At $\beta=1$
these differ from their $\lambda=0$ values
$\langle n_i n_j \rangle 
= \langle n_i \rangle \langle n_j \rangle$ = 1
only on-site, $i=j$.  That is,
pairs have formed locally, but not yet ordered between different sites.
However,  at $\beta=4$ the oscillating, and non-decaying, pattern 
indicates long range CDW formation.
Figure \ref{fig:Svsbetaall} exhibits 
the Fourier transform of Eq.\ref{eq:Scdw}, i.e. the structure factor $S_{\rm cdw}$. An additional normalization to $N=L^2$ 
is performed, so that $S_{\rm cdw}/N \propto 1/N$ at small $\beta$,
and $S_{\rm cdw}/N \propto 1$ at large $\beta$.
An abrupt change indicates the CDW transition.
The invariance of the low $T$ value across different
lattice sizes is another illustration the order 
is long-ranged.

The positions of these steps are close to the locations of the local
minima in the absolute value of the kinetic energy $K$ in 
Fig.~\ref{fig:KExvsbeta}.  We interpret this to indicate that the preferential
occupation of bands with $\epsilon({\bf k})<0$, which occurs even in the non-interacting
limit as $T$ is lowered, gets interrupted by the CDW formation.

The bottom panel of 
Fig.~\ref{fig:nnandxx}
indicates that the alternating pattern in the fermionic
density is accompanied by an alternating pattern in the phonon
displacements.  

The key features, however, of 
Fig.~\ref{fig:Svsbetaall} 
are the {\it high values of the transition temperatures} $T_{\rm cdw}$
for the larger values of $A$ where the double well phonon potential energy
favors non-zero displacements.  Typical values of $T_{\rm cdw}$ in
the conventional Holstein model are in the
range $T_{\rm cdw}/t \sim 0.2 - 0.3$ for analogous choices of $g$ and $\omega_0$
\cite{feng2020}.  In the next section, we verify that these high $T_{\rm cdw}$
persist even when the product of the electron-phonon coupling
and typical phonon displacements are restricted to be the same as in
the conventional Holstein model.

\begin{figure}[t!]
\centerline{\includegraphics[width=0.5\textwidth]{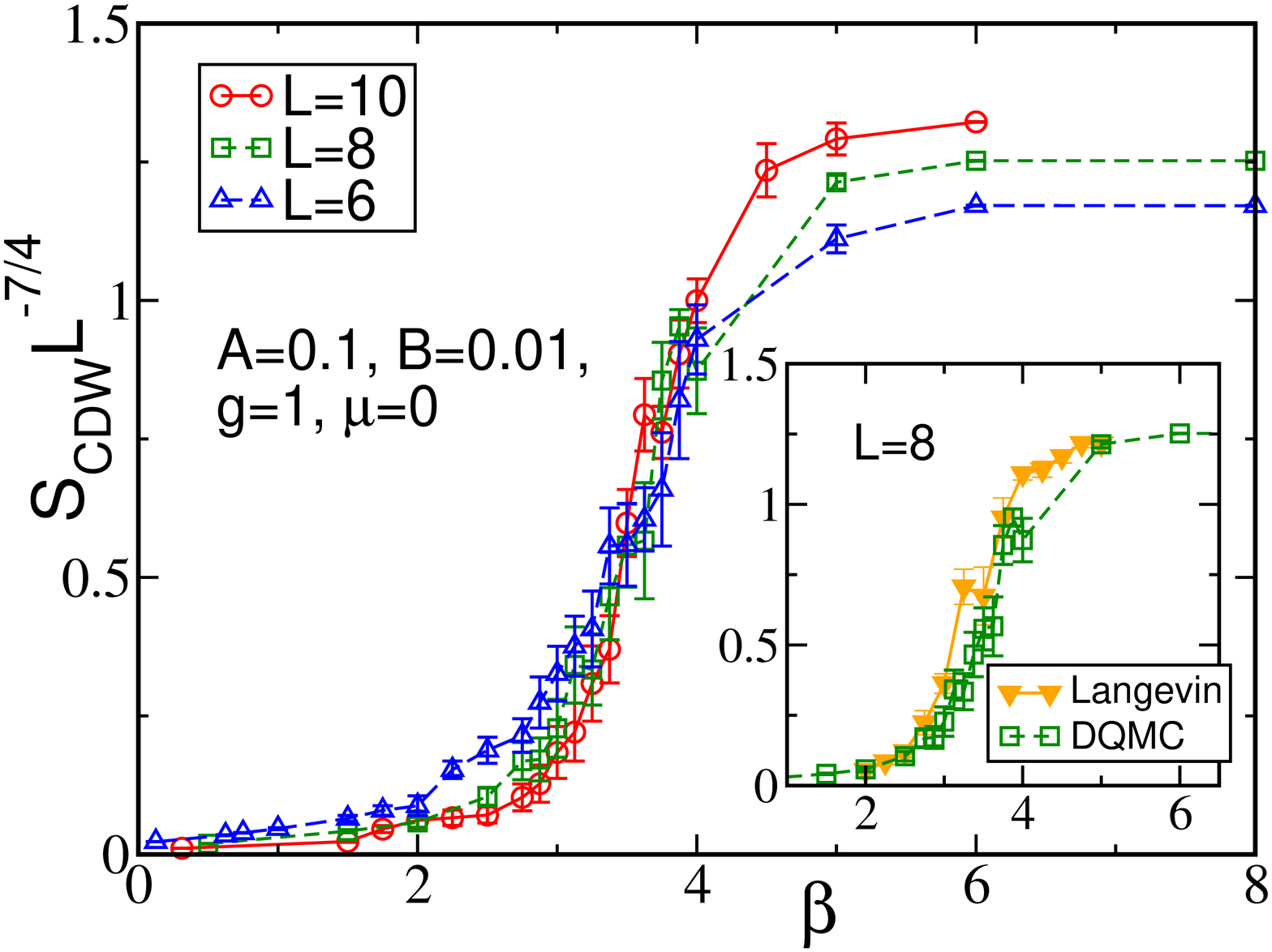}}
\includegraphics[width=0.5\textwidth]{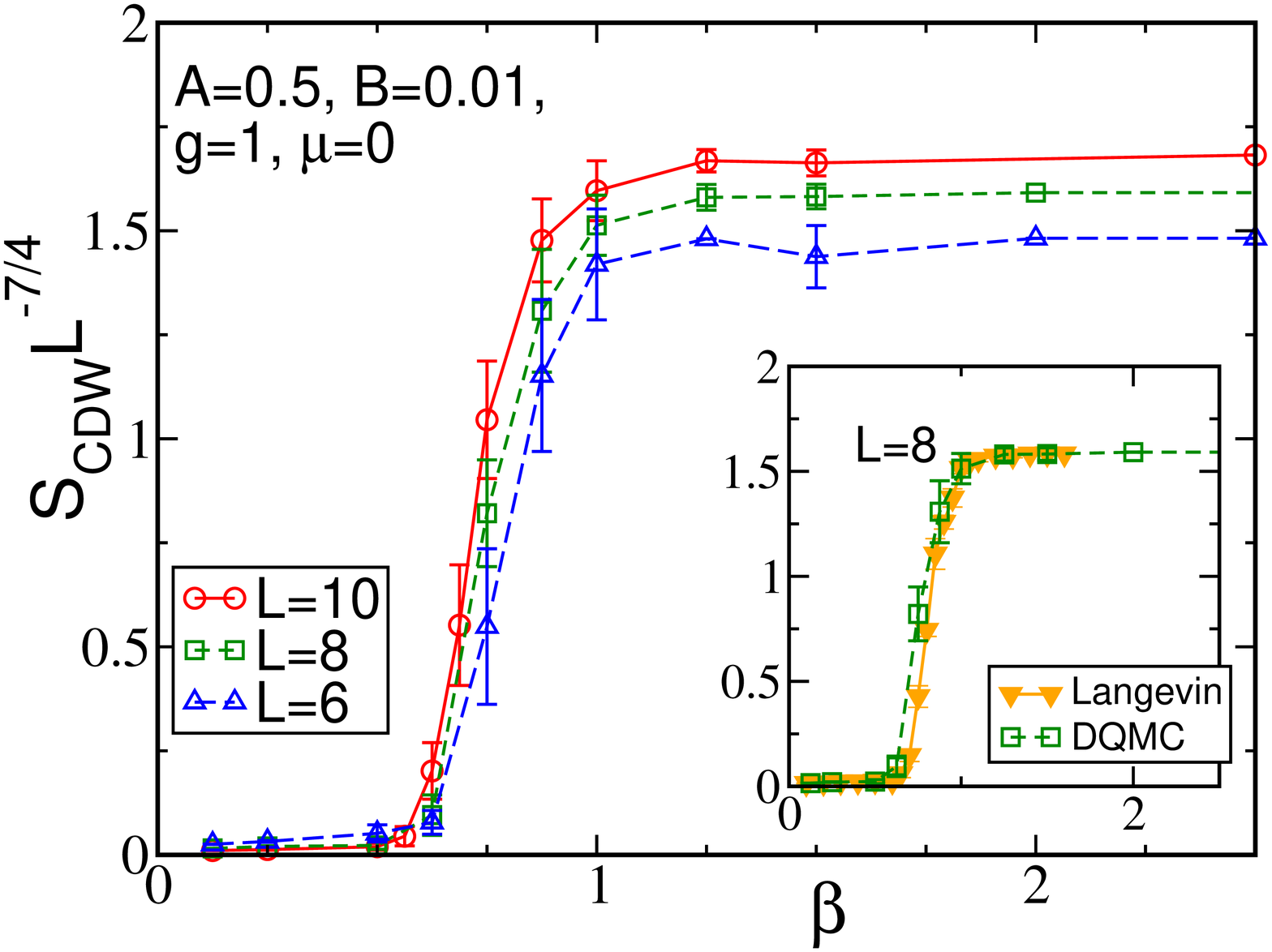}
\caption{\label{fig:SL74vsbeta}(Color online). 
Scaling analysis of the charge structure factor for $B=0.01$, $g=1$, $\mu=0$ and $A=0.1$ (top) or $A=0.5$ (bottom) with corresponding $\omega_0=\sqrt{2A}$ and $\lambda=\sqrt{2\omega_0}g$.
When $S_{\rm cdw}$ is normalized by $L^{\gamma/\nu}$
with $\gamma/\nu=7/4$, the 2D Ising values, a crossing as a function of inverse temperature $\beta$ occurs at the critical point.
 The top figure shows the $A=0.1$ case for which we observe the crossing around $\beta_{\rm cdw} = 3.5 \pm 0.2$. In the bottom figure, $A=0.5$ and $\beta_{\rm cdw} = 0.56 \pm 0.06$.
 The insets show a comparison between results
 obtained with DQMC and Langevin methods for $L=8$ in the critical region.
}
\end{figure}

We conclude this discussion by presenting a scaling analysis to determine
$T_{\rm cdw}$ more precisely. When normalized by $N^{-1}=L^{-2}$, a lattice-size independent structure factor provides evidence for ground state long range order, as already seen in Fig.~\ref{fig:Svsbetaall}.  The temperature at which this order first occurs can be determined by examining $L^{-\gamma/\nu} S_{\rm cdw}$. The theory of finite size scaling predicts that curves of  $L^{-\gamma/\nu} S_{\rm cdw}$ as functions of $T$ (or $\beta$) for different lattice sizes should all cross at one point, thus yielding the value of $T_{\rm cdw}$. Here in the Holstein model on a square lattice, the transition is in the 2D Ising universality class with $\gamma/\nu=7/4$, simplifying the analysis. Figure \ref{fig:SL74vsbeta} shows the result for the two cases with 
$A=0.1$ (top) and $A=0.5$ (bottom). The crossing is at $T_{\rm cdw} = 0.29 \pm 0.02$ ($\beta_{\rm cdw} = 3.5 \pm 0.2$) for $A=0.1$ and as high as $T_{\rm cdw} = 1.8 \pm 0.2$ ($\beta_{\rm cdw} = 0.56 \pm 0.06$) for $A=0.5$.

We also demonstrate that the two computational methods, DQMC and
Langevin, give consistent results by comparing results for $L=8$ in the insets of Fig.  \ref{fig:SL74vsbeta}.

\section{Simulations at Fixed Average Phonon Displacement}

In the preceding section we reported values for 
$T_{\rm cdw}/t$ which exceed by a factor of two or three
those obtained over a range of values of 
electron-phonon couplings $\lambda$ and 
phonon frequencies $\omega_0$
previously reported for the conventional Holstein 
Hamiltonian.  

These results are already significant
because the existence of a maximal $T_{\rm cdw}/t$
at intermediate $\lambda$ and $\omega_0$ suggests a fundamental
limit to the CDW transition temperature
in the conventional Holstein model.
However, one could still ask whether
the high critical transition temperatures 
of  Fig.~\ref{fig:Svsbetaall} 
are associated
with anomalously large phonon displacements,
or some related unphysical parameter choice.
In this section we reproduce many of the preceding
results tuning the anharmonic potential (that is, $A$ and $B$)
to keep fixed phonon displacement.
More specifically, 
we show in Appendix B that the choice
\begin{align}
A = \frac{4B x_0^3 - \lambda}{2 x_0}
\end{align}
keeps $\langle x \rangle = x_0$. Thus when we vary $A$ we do so with an accompanying change in $B$ to fix the mean phonon displacement. We chose to 
compare to the conventional Holstein model with $\lambda=2$ and $\omega_0=1$ for which $x_0=\langle x \rangle =  \lambda/\omega_0^2=2$. In addition, we use the same value of $\lambda=2$ in both models to keep the product $\lambda x$  similar. We studied two cases with $B=0.1$ and $B=0.2$ which give $A=0.3$ and $A=1.1$, respectively.

\subsection{Local Observables}

\begin{figure}[t!]
\centerline{\includegraphics[width=0.5\textwidth]{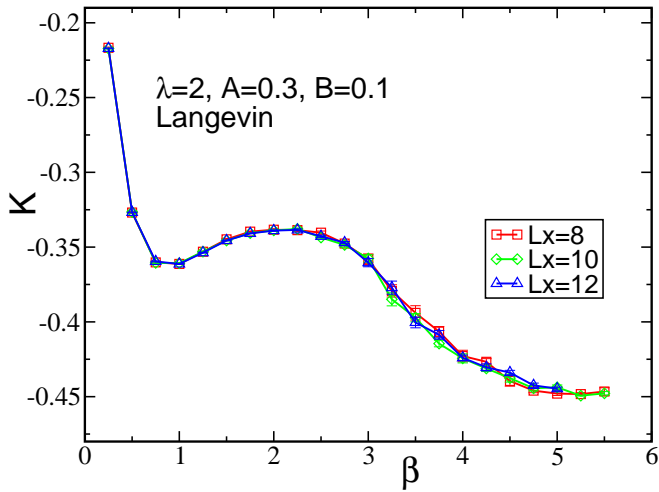}}
\caption{\label{fig:Kxvsbeta2}(Color online). 
Kinetic energy as a function of $\beta$.  
The non-monotonic behavior of the kinetic energy 
reflects the development of charge correlations.
Parameters are
$A=0.3$, $B=0.1,$ $\lambda = 2$ and have been chosen to obtain 
a phonon field $x_0=2$.
}
\end{figure}

To ensure the observation of high CDW transition 
temperatures reported in the preceding section
is robust, we focus here on measurements of long range order which
more precisely determine $T_{\rm cdw}$.  Nevertheless,
it is useful to examine one local measurement, the kinetic energy, since
its non-monotonic behavior has been seen earlier to provide an
important initial indication of
the onset of the insulating CDW phase.
Figure \ref{fig:Kxvsbeta2} exhibits this
decrease in magnitude of $K$ in the vicinity of the
CDW ordering transition.

\subsection{Long Range Charge Order}

\begin{figure}[t!]
\centerline{\includegraphics[width=0.5\textwidth]{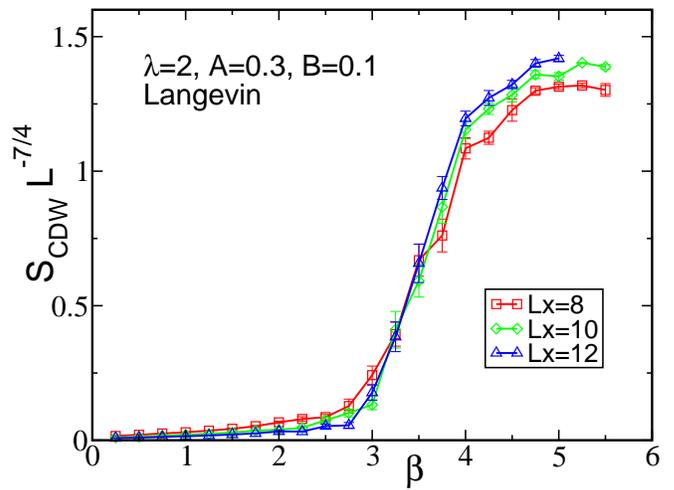}}
\caption{\label{fig:SL74vsbeta2}(Color online). Langevin data for 
$A=0.3$, $B=0.1,$ $\lambda = 2$, which corresponds to a phonon field $x_0=2$.
A crossing at $\beta_c \sim 3.25$ gives the position of the CDW transition.
}
\end{figure}

Figure \ref{fig:SL74vsbeta2} shows a finite size scaling crossing plot
for one of these `fair comparisons' in which the phonon displacement is restricted
to be the same as for the conventional Holstein model.  We find 
$\beta_{\rm cdw} \sim 3.25$ 
($T_{\rm cdw} \sim 0.31$), which is higher than the transition temperature of the Holstein model on a half-filled square lattice with $\lambda=2, \omega_0=1$ \cite{feng2020i}.
Choosing $A=1.1$ and $B=0.2$ and keeping $\lambda=2$, $\omega_0=1$ and $x_0=2$, we increase the transition temperature to $\beta_{\rm cdw}=2.5$. This shows that for the same fixed average value of lattice displacement, $x_0$, we obtain higher critical temperatures by increasing $A$ and $B$.
Furthermore, as noted earlier, $T_{\rm cdw}$ as a function of $\lambda$ in the Holstein model is non-monotonic, with a maximum $T_{\rm cdw} \sim 0.25$ at dimensionless electron-phonon coupling strength $\lambda_D \sim 0.4$ when $\omega_0=1$\cite{feng2020i}. Meanwhile, the transition temperature does not depend on $\omega_0$ sensitively as long as the effective attraction in the Holstein model $U=-\lambda^2/\omega_0^2$ is fixed \cite{zhang22}. The large $T_{\rm cdw} \sim 2$ shown in Fig.~\ref{fig:SL74vsbeta} (bottom), much higher than the maximum $T_{\rm cdw}$ we can achieve in the pure Holstein model, indicates the Holstein model with anharmonic potential we study here significantly increases the CDW phase transition temperature.
\color{black}

\bigskip
\section{Conclusions}

In this paper we have used determinant Quantum Monte Carlo and 
Langevin simulations to examine the properties of
a square lattice Holstein model with an anharmonic phonon
potential.  This potential has an intrinsic double well 
structure favoring non zero phonons fields and, consequently, empty and doubly occupied sites.
Unlike most previous extensions of the Holstein model
to include anharmonicity, our results show a marked 
increase in the CDW transition temperatures, from
$T_{\rm cdw} \sim t/6 - t/4$ for the conventional
Holstein model, to $T_{\rm cdw} \sim t/2 - t$.
Our result is not a consequence of a trivial
rescaling of $T_{\rm cdw}$ resulting from
larger phonon displacements; we demonstrated this by choosing parameter sets where the average phonon displacement is similar to those in the conventional Holstein model. In any case, in the Holstien-Hubbard model, $T_{\rm cdw}$ has a maximum as a function
of electron-phonon coupling, phonon frequency, and the
resulting phonon displacement, which is well below
the transition temperatures found here.

It would be interesting to explore superconducting correlations
in this model.  One expects CDW and SC to be competitive, so that
the emergence of SC will surely require doping away from half-filling.
QMC is especially useful here, since it will allow comparison
to analytic approaches like Migdal-Eliashberg theory\cite{migdal58,eliashberg60} which have been critical to the understanding of the conventional Holstein
Hamiltonian\cite{alexandrov01,bauer11,esterlis18,dee20},
and which have been extended to include anharmonicity
\cite{hui74,kavakozov78,mahan93}.

Double well phonon potentials such as we consider, have been suggested
to provide a counterpart to the Kondo effect\cite{fuse12}.  There, electrons interact with local spin-1/2 degrees of freedom, resulting in counter-intuitive transport properties like the resistance minimum of heavy fermion systems\cite{stewart84,gegenwart08}. In the phonon case, nuclei can analogously `rattle' between two minima, and via an appropriate coupling to electrons renormalize their mass.

Furthering this similarity, in the dense limit of the 
Kondo Lattice (KL) model\cite{auerbach86,tsunetsugu97,steglich16}, the spins can order antiferromagnetically via an indirect Ruderman-Kittel-Kasuya-Yosida (RKKY)\cite{ruderman54,kasuya56,yosida57} interaction mediated by conduction electrons. Although we have mainly characterized our CDW phase as one in which the electron density is modulated, there is an accompanying alternation of phonon coordinates in our model, as seen in  the bottom panel of Fig.~\ref{fig:nnandxx}. Since our phonon degrees of freedom are not directly coupled to each other, this oscillating structure forms via coupling to the conduction electrons.

Indeed, this sort of analogy to Kondo physics in the
context of phonons has been considered by 
Anderson and Yu \cite{yu84} to explain properties like the 
high SC transition temperatures,
and resistivity saturation of A-15 materials.

\begin{acknowledgments}
The work of RTS was supported by the 
grant DE‐SC0014671 funded by
the U.S. Department of Energy, Office of Science.
CK's work was supported by the UC Davis Physics REU program under NSF grant PHY2150515.
\end{acknowledgments}

\newpage

\vskip0.10in \noindent
{\bf Appendix 1:   
Particle-Hole Symmetry
in the presence of an Anharmonic Potential}

\vskip0.10in 
There are two related ways to discuss the particle-hole symmetry of the model.  The first is to
consider a single site model ($t=0$)  with the phonon potential of Eq.~\ref{eq:ham},
\begin{eqnarray}
V(x) = \frac{1}{2}\omega_0 ^2 x^2 + \lambda x n - \mu n.
\end{eqnarray}

The density is given by,
\begin{align}
\label{n_up}
\langle n_\uparrow \rangle &= Z^{-1} \sum_{{n}_{i\uparrow}=0}^{1} \sum_{{n}_{i\downarrow}=0}^{1} \int dx \, n_\uparrow \, e^{-\beta V(x)} 
\nonumber \\
Z &=  \sum_{{n}_{i\uparrow}=0}^{1} \sum_{{n}_{i\downarrow}=0}^{1} \int dx \, e^{-\beta V(x)} 
\end{align} 
If we introduce the notation $I(n_\uparrow, n_\downarrow)$ to denote the integral for a specific choice of number operators, we can re-write Eqn. \ref{n_up} as:
\begin{eqnarray}
\label{n_up_int}
\langle n_\uparrow \rangle = \frac{I(1,0) +I(1,1)}{I(0,0) + 2I(1,0) + I(1,1)} 
\end{eqnarray}
where the denominator is the partition function. Rearranging this shows that the half-filling
condition $\langle n_\uparrow \rangle = 1/2$ is $I(0,0)=I(1,1)$, which can only be
 true when $\mu = 0$. 
When $\mu = 0$, the curves of $V(x)$ for $n=0$ and
 $n=2$ are reflections of each other in the y axis, thus giving us symmetry between 
the ``hole" and ``particle" curves.  

A more formal analysis is to 
apply a particle-hole transformation (PHT), $d_{i \sigma}^{\phantom{\dagger}}=(-1)^{i}c_{i \sigma}^{\dagger}$, 
on the Hamiltonian.  Here $(-1)^i$ means a phase of $-1$ on one sublattice
and $+1$ on the other sublattice of the bipartite square geometry. This choice of phase ensures the electron hopping term remains the same 
under the PHT.  Meanwhile, the density operator $n_{i\sigma}$ transforms
into $1-n_{i\sigma}$.  If we also introduce $y_i = -x_i$ we see that the original Hamiltonian
is recovered  except for a change in sign of the
chemical potential $\mu$.  
 This demonstrates that 
density of the system obeys $n(\mu)=2-n(-\mu)$.  From this, it is obvious that $\mu=0$ yields half-filling $n=
\langle n_\uparrow + n\downarrow \rangle = 1$.

\vskip0.10in \noindent
{\bf Appendix 2:  Relation between $A$ and $B$ to fix $x_0$}

\vskip0.05in \noindent
In order to compare results of simulations of the anharmonic model to the original Holstein Hamiltonian, 
setting the el-ph coupling $\lambda$ and phonon frequency $\omega_0$ 
(with $\omega_0 = \sqrt{2A}$) to be the same, as done in Sec.~III, is not
sufficient.  The reason is that the electrons move in an energy landscape given by the {\it product} of
$\lambda$ and phonon displacement.  
A comparison which ensures equivalence of the
energy landscape is obtained by requiring that
$\lambda x_0$ be the same in the double well potential as
in the conventional Holstein model.  Here $x_0$ is  the position of the minima in the phonon potential
corresponding to empty ($n=0$) and doubly occupied ($n=2$) sites.

In the conventional Holstein Hamiltonian, at half-filling ($\mu=0$)
\begin{align}
V = \frac{1}{2}\omega_0 ^2 x^2 - \lambda x (n-1) .
\end{align}
and the minima are at 
$x_0 = \pm \lambda/\omega_0^2$ for $n=2$ and $n=0$, respectively. 
It is straightforward to determine $A,B$ in the  
anharmonic double well potential to give the same $x_0$.
The phonon potential is
\begin{align}
V = -Ax^2 + Bx^4 - \lambda x (n - 1).
\end{align}
with half filling again at $\mu = 0$.  The minima of the $n = 2$ 
curve is at positive $x_0$ (the minima for $n=0$ being at $-x_0$) and given by the condition
\begin{align}
-2 A x_0 + 4 B x_0^3 - \lambda = 0.
\end{align}
Therefore, to keep the location of the minima fixed, 
$A$ and $B$ must satisfy 
\begin{equation}
A = \frac{4B x_0^3 - \lambda}{2 x_0}.
\label{eq:ABreln}
\end{equation}
In addition, one should use the same value
of $\lambda$ in both models so that the product $\lambda x_0$ is the same.
Thus in Sec.~IV
we proceed by fixing a (small) $B$ and using 
Eq.~\ref{eq:ABreln} to determine $A$.
Commonly used parameters are, for example, $\lambda=2$ and $\omega_0=1$ which yield $x_0=2$. We used these parameters for comparison.
We note that the
height of the barrier 
at $x_0$  
between the minima is given by 
$Ax_0^2 + B x_0^4 - \lambda x_0$. 

\vskip0.10in \noindent

\bibliography{kvande}

\end{document}